\documentclass[conference]{IEEEtran}
\IEEEoverridecommandlockouts
\usepackage{cite}
\usepackage{mathtools,cuted}
\usepackage{amsmath,amssymb,amsfonts,hyperref}
\hypersetup{colorlinks = true,linkcolor= blue,citecolor=red}
\usepackage{algorithmic}
\usepackage{caption,subcaption}
\usepackage{graphicx}
\usepackage{textcomp}
\usepackage{xcolor}
\usepackage{multirow}
\usepackage{lipsum}
\usepackage{lscape}

\def\BibTeX{{\rm B\kern-.05em{\sc i\kern-.025em b}\kern-.08em
		T\kern-.1667em\lower.7ex\hbox{E}\kern-.125emX}}
\begin{document}
	\title{Seeing is Believing: A Federated Learning Based Prototype to Detect Wireless Injection Attacks\\
	}
 \author{\IEEEauthorblockN{Aadil Hussain$^{*}$, Nitheesh Gundapu$^{*}$, Sarang Drugkar$^{*}$, Suraj Kiran$^{*}$, J. Harshan$^{*}$, Ranjitha Prasad$^{\dagger}$}
\IEEEauthorblockA{$^{*}$Indian Institute of Technology Delhi, India, $^{\dagger}$Indraprasta Institute of Information Technology, India}
}

	\maketitle
\begin{abstract}
Reactive injection attacks are a class of security threats in wireless networks wherein adversaries opportunistically inject spoofing packets in the frequency band of a client thereby forcing the base-station to deploy \emph{impersonation-detection} methods. Towards circumventing such threats, we implement secret-key based physical-layer signalling methods at the clients which allow the base-stations to deploy machine learning (ML) models on their in-phase and quadrature samples at the baseband for attack detection. Using Adalm Pluto based software defined radios to implement the secret-key based signalling methods, we show that robust ML models can be designed at the base-stations. However, we also point out that, in practice, insufficient availability of training datasets at the base-stations can make these methods ineffective. Thus, we use a federated learning framework in the backhaul network, wherein a group of base-stations that need to protect their clients against reactive injection threats collaborate to refine their ML models by ensuring privacy on their datasets. Using a network of XBee devices to implement the backhaul network, experimental results on our federated learning setup shows significant enhancements in the detection accuracy, thus presenting wireless security as an excellent use-case for federated learning in 6G networks and beyond.
     \end{abstract}


\section{Introduction}

Owing to the widespread use of wireless communication, it is crucial to secure next-generation of wireless networks from a wide range of active threats including malicious impersonations, injection attacks, etc on its wireless clients \cite{a30}. Although defenses in the form of software attestation, anomaly detection, trust-based and signature-based techniques can be implemented on their base-stations \cite{a29}, \cite{a2}, \cite{a4}, \cite{s110}, these mechanisms often come with limitations \cite{a30}, thus requiring new robust solutions. In the recent years, machine learning (ML) techniques have demonstrated exceptional capabilities in pattern recognition, making them a strong toolset for distinguishing adversarial and non-adversarial entities in the wireless world \cite{a5}. In such techniques, the underlying ML algorithms perform data-driven learning, wherein they identify and learn the relevant features from a given dataset of observations. Subsequently, the trained neural networks are deployed at the base-stations to distinguish adversarial and non-adversarial entities using their observations. 

Although base-stations are computationally capable of hosting ML techniques to detect impersonation attacks on their clients, they may not witness sufficient ``labelled" datasets to train their neural networks. As a consequence, the ML based detection technique, when deployed at the base-station in the run-time, may not provide high-accuracy in their classification performance. One way to circumvent this problem is to involve a cloud-centric architecture, wherein the labelled datasets at each base-station are communicated to a cloud infrastructure, and an ML based neural network is trained at the cloud using the union of the datasets from all the base-stations. While this solution may enhance the detection accuracy, it accompanies the following concerns: (i) if the participating base-stations belong to different (or competing) service providers, they may not be willing to share the datasets of their clients due to privacy concerns, (ii) even if the participating base-stations belong to the same service provider, they may need to provide confidentiality feature on their datasets through crypto-primitives, which in-turn increases communication and complexity overheads, (iii) finally, when the ML model is implemented at the cloud, the observations that are under scrutiny for detection at each base-station, must be communicated to the cloud, thereby increasing the latency of detection \cite{s1}. 

Owing to the drawbacks of a centralized architecture for ML models, it is appealing to retain the labelled datasets locally at each base-station, and still train their ML model in such a way that the accuracy benefits of that of the central architecture can be approached. One such decentralised strategy is the framework of Federated Learning (FL), wherein the participating base-stations collaborate to train a single model, by exchanging only their locally calculated updates with a centralized sever, thereby preserving privacy \cite{s3}. A popular FL implementation is federated averaging \cite{s2}, which iteratively computes the local models at each base-station, and subsequently incorporates them into a global model via communication with a server. While FL seems to be an appealing choice to implement decentralized ML techniques, identifying strong use-cases in wireless networks and studying their implementation aspects are imperative. In this contribution, we present an interesting security use-case in wireless networks, wherein base-stations of different networks benefit from implementing FL based solution for detecting injection attacks on their clients. Explicit details on our contributions are provided in the next section.        

\begin{figure*}[ht!]
		\centering
		\includegraphics[scale=0.43]{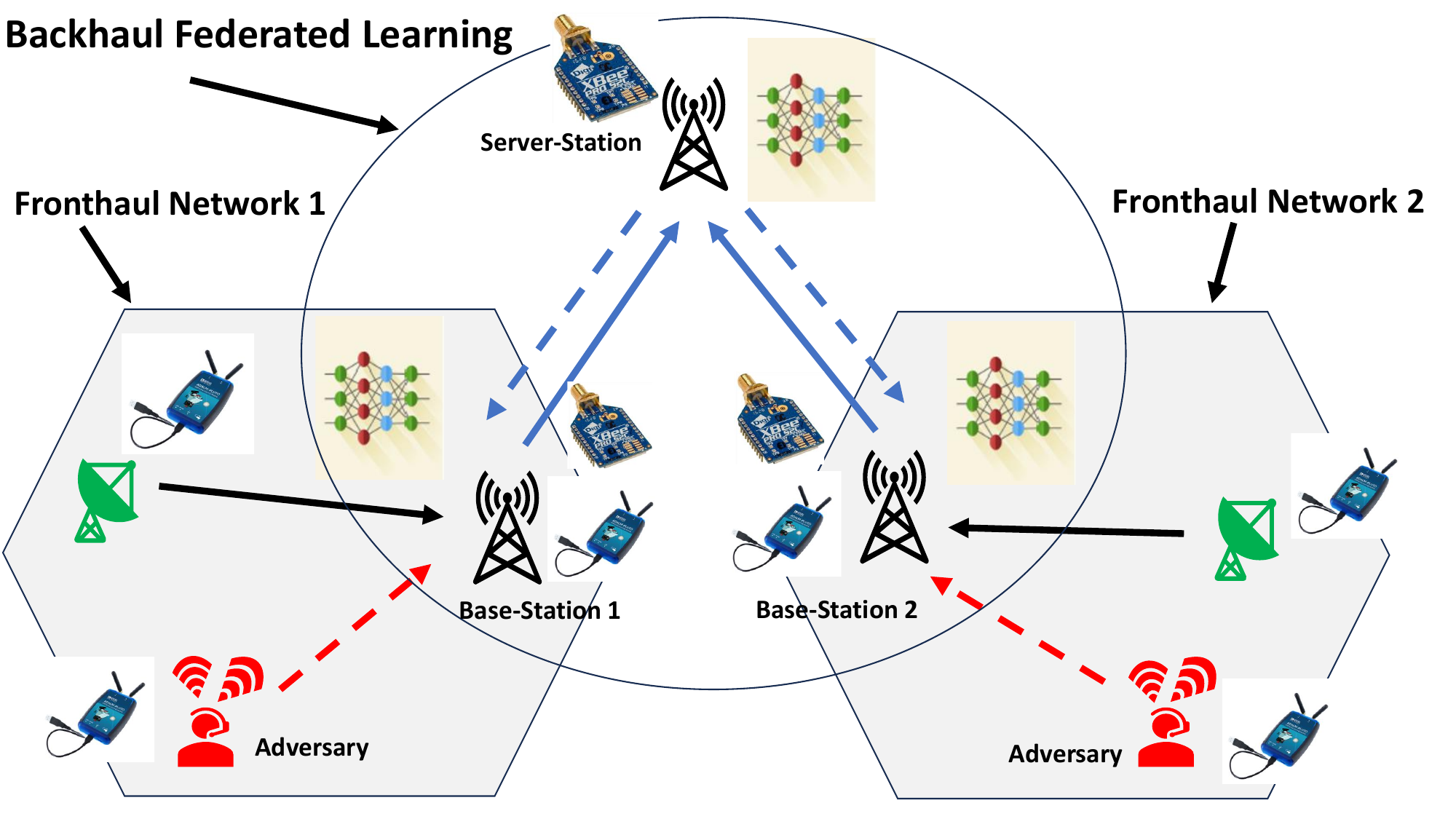}
  \vspace{-1mm}
		\caption{In our setup, the base-stations of individual networks implement learning models to detect inject attacks in the fronthaul in their cell. They also refine their neural networks in the backhaul through federated learning among multiple base-stations. In our prototype, we have used Adalm Pluto SDRs for fronthaul and XBee devices for backhaul operations.}
		\label{model}
	\end{figure*}


\subsection{Contributions}

As depicted in Fig. \ref{model}, we propose a wireless security use-case wherein base-stations of individual networks are interested in mitigating a common class of injection attacks on their clients. To detect these threats, we first present a secret-key based mitigation strategy at the victims (see Section \ref{SM}), and then show that Convolutional Neural Networks (CNN) can be used at the base-stations to classify the received frames as adversarial or otherwise (see Section \ref{identify}). Once a CNN based detection is formalized at each base-station, we leverage on the benefits of FL by asking the base-stations to  collaborate and refine their neural models. For the prototype demonstration of this use-case, we use Adalm Pluto based software defined radios (SDRs) at the clients and a combination of Adalm Pluto SDRs and XBee devices at the base-stations of individual networks. The interface is done in such way that each base-station uses SDRs in the fronthaul, whereas XBee devices are used in the backhaul communication to implement FL (see Section \ref{fl_inject}). Experimental results from our prototype reveal that FL assists in significant improvements in the accuracy of the proposed detection method. Overall, this work is the first of its kind to demonstrate the benefits of FL in detecting injection attacks in wireless networks. 
 
\section{System Model}\label{SM}

\begin{figure}[ht!]
		\centering
		\includegraphics[scale=0.36]{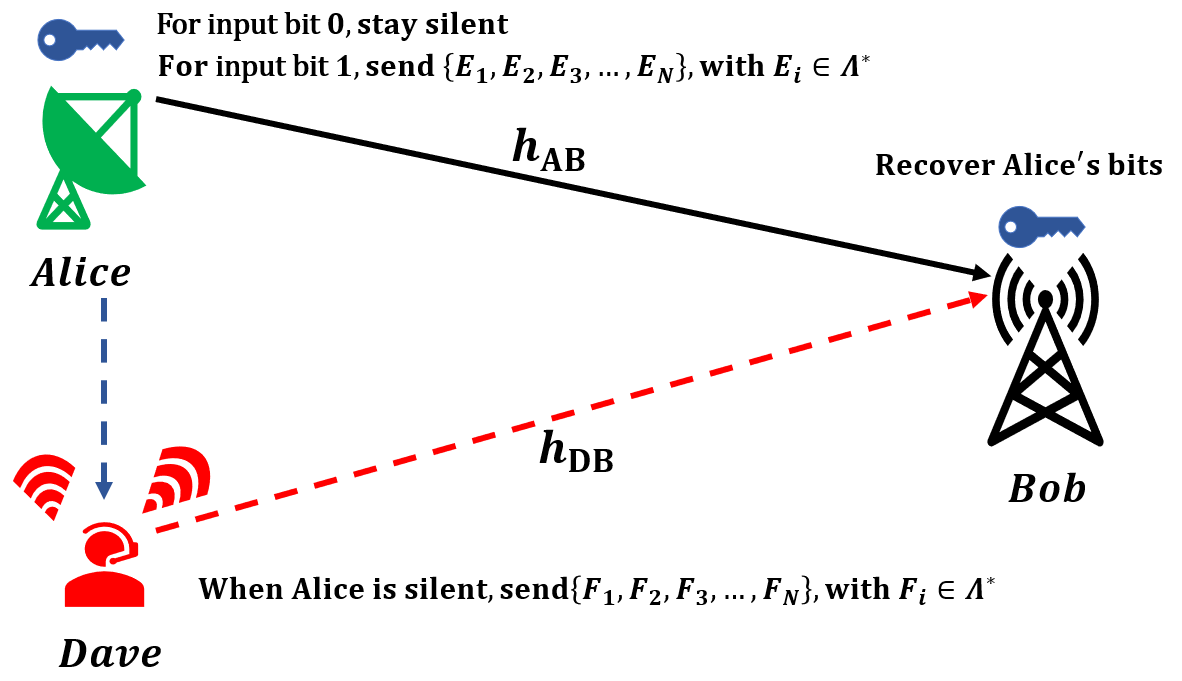}
		\caption{Depiction of a countermeasure to the injection attack on Alice's OOK symbols. Instead of fixed energy levels for bit-1, a random sequence of energy levels are chosen from a dictionary at Alice using a secret-key with Bob.}
		\label{sigmod}
	\end{figure}

In this section, we first present the single base-station based adversarial network by covering the signalling methods employed at the legitimate nodes and the adversary. Consider two legitimate nodes, Alice and Bob, as illustrated in Fig. \ref{sigmod}, that play the roles of a client and a base-station (BS), respectively. We assume that Alice uses On-Off Keying (OOK) to convey her information bits to Bob. The reasons for opting OOK are two-fold: (i) to minimize pilot-transmission overhead from coherent communication and (ii) to avoid a potential pilot-contamination attack by an active adversary in the network. We also assume the presence of an adversarial node named Dave, that intends to execute an injection attack on the communication between Alice and Bob. In particular, Dave is equipped with a perfect full-duplex radio to eavesdrop on Alice's transmitted bits. Furthermore, Dave has the capability to inject non-zero energy levels on Alice's communication band whenever Alice communicates bit-0, and remain silent when bit-$1$ is transmitted. As a consequence, when the attack is perfectly implemented, non-zero energy levels are ensured on all the bit-patterns of Alice, which in turn results in an overall error rate of $50\%$ at Bob.

\subsection{Secret-Key Based Mitigation Strategy}\label{key_model}
 
To mitigate the above injection attack, we employ a secret-key based signalling scheme as presented in \cite{s4}. In such a scheme, the signal transmission unfolds as follows: Alice utilizes a repetition-coding strategy to communicate a single bit to Bob in $N$ channel-uses, for some $N > 0$. First, Alice and Bob agree upon a dictionary of energy levels, denoted by $\mathbf{\Lambda}^{*} = \{\delta_{1}, \delta_{2}, \ldots, \delta_{L}\}$ such that $\delta_{j} > 0$ for each $j \in [L]$. Subsequently, for bit-$1$, Alice generates $N$ independent energy levels, denoted by $\mathbf{s} = [E_{1}, E_{2}, \ldots, E_{N}]$ from the dictionary $\mathbf{\Lambda}^{*}$. Note that this pseudo-random choice of the $N$ energy levels are based on secret-key already shared with Bob. Conversely, for bit-$0$, Alice remains silent, i.e., she transmits an all-zero vector of length $N$. Based on the threat model, Dave, which is familiar with the coding scheme and the dictionary, remains silent upon detecting bit-$1$ from Alice. However, when he detects bit-$0$ from Alice, Dave randomly selects $N$ independent energy levels, denoted by $\mathbf{w} = [F_{1}, F_{2}, \ldots, F_{N}]$ from the dictionary, and then transmits them to Bob since he does not have the knowledge of the secret-key. Thus, when bit-$1$ is sent by Alice, the received baseband symbols at Bob across the $N$ time slots are of the form
\begin{equation}	
\mathbf{y}=\mathbf{h}_{AB} \odot \sqrt{\textbf{s}} +\mathbf{z}_{B},
 \label{alice}
\end{equation}
where $\mathbf{h}_{AB}\in \mathbb{C}^{1 \times N}$ is the channel vector from Alice to Bob such that each component is i.i.d. with $\mathcal{CN}(0,1)$, $\mathbf{z}_{B} \in \mathbb{C}^{1 \times N}$ is the Additive White Gaussian Noise (AWGN) such that each component is i.i.d. with $\mathcal{CN}(0,N_{0})$ for some $N_{0} > 0$. Finally, $\sqrt{\textbf{s}}$ denotes the vector such that $\sqrt{.}$ is applied on each component of $\textbf{s}$. Also, $\odot$ denotes the component-wise product of two vectors. 
 
Similarly, when Alice transmits the all-zero codeword, the adversary detects it and sends an i.i.d. energy sequence $\textbf{w}$ with components coming from $\mathbf{\Lambda}^{*}$. Consequently, the received baseband symbols at Bob, involving $N$ time slots are of the form
\begin{equation}
\mathbf{y}=\mathbf{h}_{DB} \odot \sqrt{\textbf{w}} +\mathbf{z}_{B},
\label{dave}
\end{equation}
where $\mathbf{h}_{DB}\in \mathbb{C}^{1 \times N}$ is the channel vector between Dave and Bob. Given a received vector $\textbf{y}$, the main objective of the detection strategy at Bob is to classify whether $\mathbf{y}$ is of the form \eqref{alice} or \eqref{dave} without the knowledge of the channel, i.e., to detect whether bit-$1$ or bit-$0$ was sent by Alice. Since Alice chooses an energy sequence from a set of $L^{N}$ sequences based on a secret-key, the adversary will be able to match this sequence in ideal conditions with probability $\frac{1}{L^{N}}$, which is negligible as long as $N$ is sufficiently large. Given that the vector $\mathbf{s}$ is known to Bob under the hypothesis that bit-1 is sent by Alice, a maximum a posteriori (MAP) detector can be formulated at Bob using the statistical distributions of $\mathbf{h}_{AB}$ and $\mathbf{h}_{DB}$. However, since the channel $\mathbf{h}_{DB}$ between Dave and Bob, its distribution is challenging to obtain in practice, and therefore, using a MAP detector for detection is not realistic.

To overcome the above mentioned problem with MAP detector, in the next section, we propose a learning based detection method to classify whether $\mathbf{y}$ is of the form \eqref{alice} or \eqref{dave} using \emph{a priori} obtained datasets. First, we discuss the idea of how a CNN based classifier can be employed on the received vector $\mathbf{y}$, and then discuss the implementation aspects of our prototype.   

 \section{Key-assisted CNN based Classifier}\label{identify}
 

Having received $\mathbf{y}$, Bob now needs to classify whether the transmitted sequence was from Alice or otherwise in order to decode her information bit. Since Bob has the key, he uses it to locally generate the sequence of energy levels $\mathbf{s}$ that would be sent if Alice had bit-$1$ during that frame. As a result, Bob modifies $\mathbf{y}$ to obtain
\begin{equation}
\tilde{\mathbf{y}} = \mathbf{y} \odot \sqrt{\mathbf{s}}^{-1},
\end{equation}
where each component of $\sqrt{\mathbf{s}}^{-1}$ is the multiplicative inverse of the corresponding component of $\sqrt{\mathbf{s}}$. Thus, under the hypotheses that Alice had sent bit-$0$ and bit-$1$, the corresponding $\tilde{\mathbf{y}}$ would be denoted by $\mathbf{\tilde{y}_w}$ and $\mathbf{\tilde{y}_s}$, respectively. In particular, we have  
\begin{equation}
\mathbf{\tilde{y}_s} = \mathbf{h}_{AB} + \mathbf{z}_{B} \odot \sqrt{\mathbf{s}}^{-1},
\end{equation}
\begin{equation}
\mathbf{\tilde{y}_w}= \mathbf{h}_{DB} \odot \sqrt{\mathbf{w}} \odot \sqrt{\mathbf{s}}^{-1} + \mathbf{z}_{B} \odot \sqrt{\mathbf{s}}^{-1}.
\end{equation}
Evidently, the structure of the symbols on $\mathbf{\tilde{y}_s}$ and $\mathbf{\tilde{y}_w}$ are distinct, and this could be used by Bob to detect an injection attack on that frame. To give an intuition on this observation, note that at high signal-to-noise-ratio values, the components of $\mathbf{z}_{B} \odot \sqrt{\mathbf{s}}^{-1}$ can be neglected, and therefore, the statistical distribution on $\mathbf{\tilde{y}_s}$ is approximately same as that on $\mathbf{h}_{AB}$. Along the similar lines, the statistical distribution on  $\mathbf{\tilde{y}_w}$ is significantly different owing to the presence of $\mathbf{w}$ and $\mathbf{s}$ in it. Given that the statistical distribution on $\mathbf{h}_{AB}$ is known (owing to the channel from a legitimate node) at Bob, we use this information to extract a set of features to classify whether $\mathbf{y}$ is of the form  \eqref{alice} or \eqref{dave}. 

\begin{figure}[ht!]
  \begin{subfigure}[b]{0.2\textwidth}
     \centering
     \includegraphics[width=\textwidth]{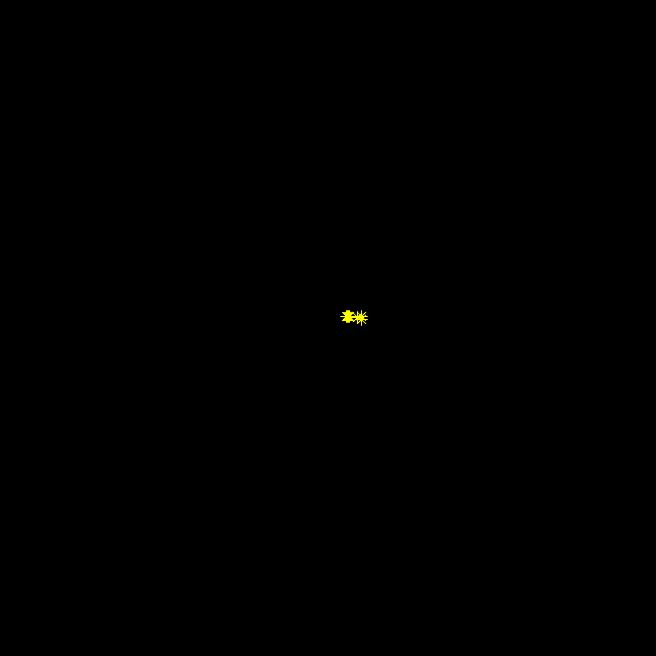}
    \caption{Alice's frame}
    \label{cd2}
 \end{subfigure}
  \hfill
 \begin{subfigure}[b]{0.2\textwidth}
     \centering
     \includegraphics[width=\textwidth]{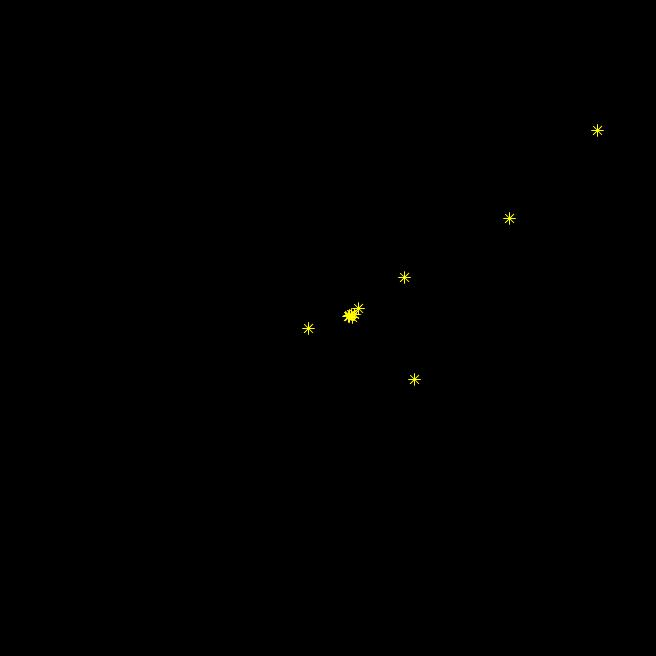}
    \caption{Dave's frame}
    \label{cd}
 \end{subfigure}
  \caption{Scatter plots of the I/Q samples of $\tilde{\mathbf{y}}$ when the frame is sent by Alice (plot on the left-hand-side) and Dave (plot on the right-hand-side). Underlying parameters for the plots are $N = 20$, $L = 8$ and SNR = $30$ dB.}
  \label{Decoded_signals}
\end{figure}

\begin{figure}[ht!]
		\centering
		\includegraphics[scale=0.32]{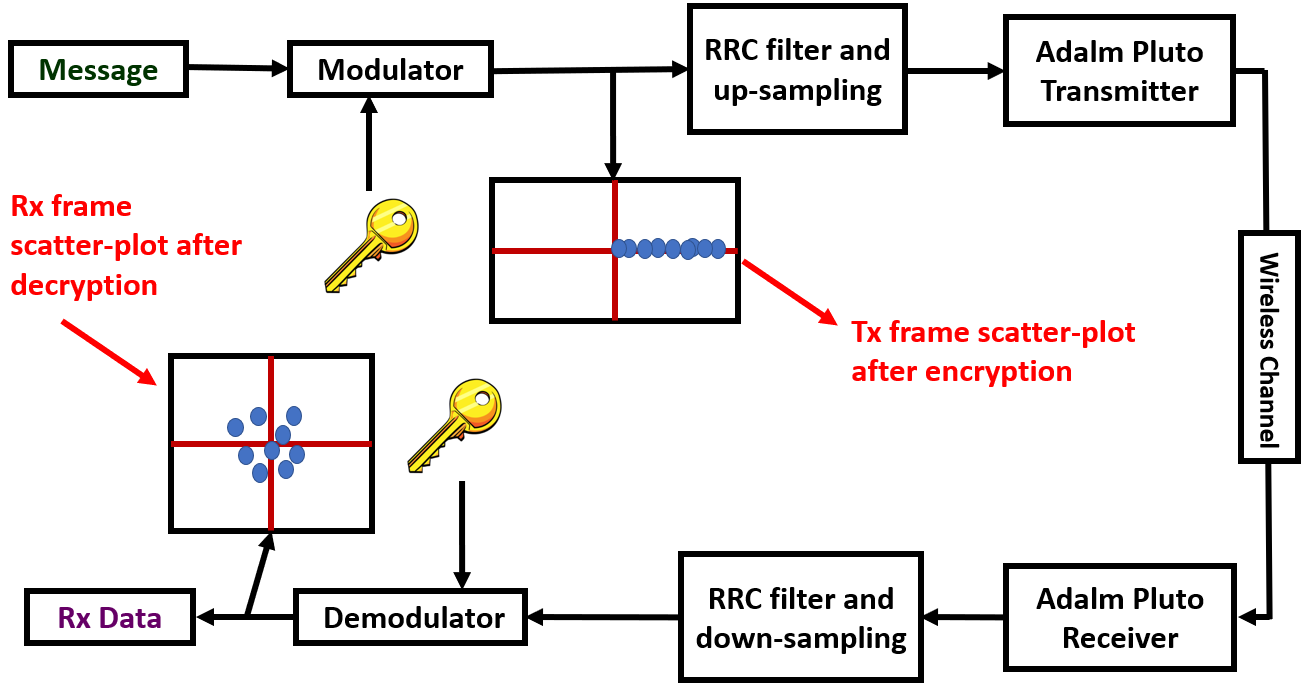}
		\caption{Proposed detection strategy wherein scatter plots are used at the base-stations to distinguish Alice and Dave. Use of secret-keys gives rise to different patterns in scatter plots.}
		\label{rff}
\end{figure}

Prior to extracting the relevant features from $\tilde{\mathbf{y}}$, we observe the scatter plots of the in-phase and quadrature (I/Q) samples of $\tilde{\mathbf{y}}$, under both hypotheses. For instance, in Fig. \ref{Decoded_signals}, we present one such plot for $\tilde{\mathbf{y}}_{s}$ and $\tilde{\mathbf{y}}_{w}$ when using the parameters $N = 20$, $L = 8$ at SNR of $30$ dB at Bob. The figures show that the scatter plot from Alice's frame appear around a confined zone as they are encoded and decoded using the same pre-shared key. Conversely, the scatter plot from Dave's frame showcases a much diverse spread as they are encoded and decoded using different keys. With this observation on a wide range of scatter plots for various values of $L$, $N$ and SNR, we resort to using the scatter plots on every frame to perform the classification task depending on the degree of the spread of the I/Q samples (as shown in Fig. \ref{rff}). Precisely, we propose to use the well-known CNN models to train a neural network at Bob using datasets that are none other than a set of vectors of the form $\tilde{\mathbf{y}}_{s}$ and $\tilde{\mathbf{y}}_{w}$ for both the classes. Subsequently, we propose to deploy this trained network at Bob to classify the frames in run-time, in order to decode Alice's bits.  

In the subsequent sections, we present a prototype of software defined radios, wherein we explain the hardware setup, and then present details on the CNN architecture. 

\subsection{Testbed Setup using Software Defined Radios}
\label{cnn_testbed}

\begin{figure}[ht!]
  \vspace{-4mm}
    \centering
    \includegraphics[scale=0.3]{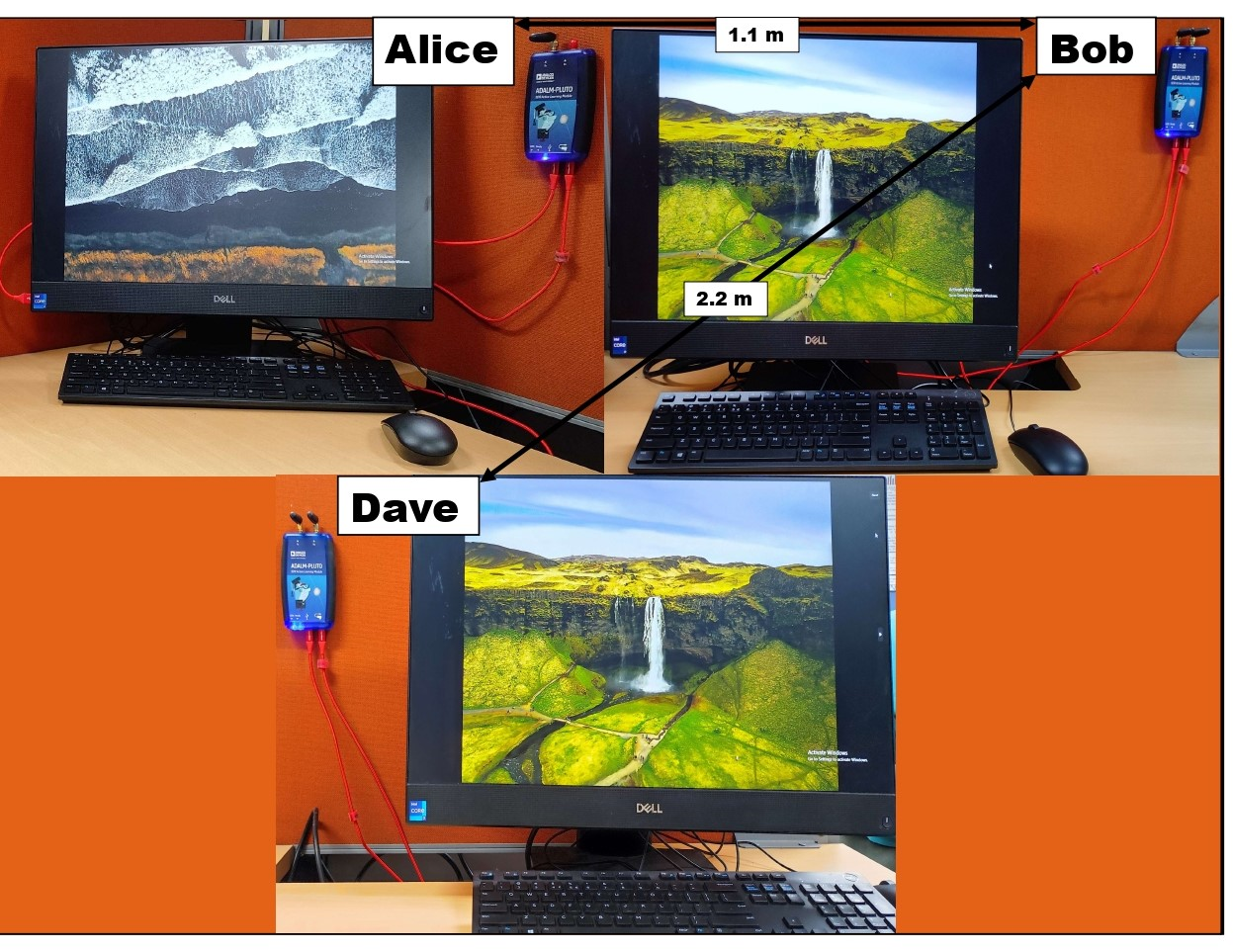}
    \caption{Depiction of a prototype to demonstrate wireless injection attacks using Adalm Pluto SDRs. This setup is used to collect I/Q samples as datasets for CNN based classifier.}
    \label{adalm}
\end{figure}

As shown in Fig. \ref{adalm}, Adalm Pluto devices are used to play the roles of Alice, Bob and Dave, wherein their baseband signal processing are implemented using MATLAB and Simulink interfaces on a computer. By using the hardware and software details as provided in Table \ref{front-table}, the following steps were executed. PlutoSDR package was installed along with the required dependencies on the computers designated for baseband signal processing. Subsequently, Adalm Pluto SDR was connected to the computer. Then, PlutoSDR flowgraph was designed for data transmission, wherein MATLAB/Simulink was used to generate a frame via the Adalm Pluto  Transmitter system object. Finally, PlutoSDR flowgraph was designed for the collection of I/Q samples at the receiver using the Adalm Pluto Receiver system object.
\begin{table}[!ht]
\centering
\addtolength{\tabcolsep}{-14.5pt}
\caption{\small{\textbf{Hardware and Software Details for SDRs}}}
\label{front-table}
\resizebox{7.5cm}{!}{%
\begin{tabular}{|c|c|}
\hline
\textbf{Hardware} & \textbf{Software} \\ \hline
\multirow{3}{*}{$3$ Adalm Pluto SDR devices.} & Communications Toolbox Support\\
&  Package for Analog Devices\\ & ADALM-Pluto Radio\\ \hline
\multirow{2}{*}{$3$ Windows 11 Systems (i$7$ $9$th gen).} 
                & Python and MATLAB \\ \cline{2-2}
                & SDR drivers and libraries.\\ \hline
\end{tabular}%
}
\vspace{-4mm}
\end{table} \\

\subsection{Implementation of the Mitigation Strategy}
\label{RFF_model}

To implement the mitigation strategy in our testbed, Alice uses the dictionary $\mathbf{\Lambda}^{*}$ = \{0.0010, 0.0055, 0.0204, 0.0690, 0.2295, 0.7703, 2.7315, 12.1727\}, with $L = 8$. Choosing the frame length $N = 20$, during the encoding process, bit-$1$ is mapped to an i.i.d. sequence from $\mathbf{\Lambda}^{*}$ based on a shared secret-key with Bob. This baseband sequence, which is passed through the transmitter root raised cosine (RRC) filter, undergoes up-conversion, amplification, and is then transmitted via Adalm Pluto transmitter. At Bob, the transmitted signal is captured using the Adalm Pluto receiver, where it undergoes down-conversion through the receiver RRC filter and is then decoded using the secret-key to obtain $\tilde{\mathbf{y}}_{s}$. For bit-$1$, with several realizations of the secret-key at Alice, an ensemble of $\tilde{\mathbf{y}}_{s}$ are collected and stored at Bob. In order to mimic the attack when bit-$0$ is transmitted by Alice, Dave's Adalm Pluto, which is placed at a different location in the lab, transmits an i.i.d. sequence from $\mathbf{\Lambda}^{*}$ similar to that of Alice, however, without using the secret-key. The corresponding set of received frames of the form $\tilde{\mathbf{y}}_{w}$ are collected and stored at Bob. The next section describes how the ensembles of $\{\tilde{\mathbf{y}}_{s}\}$ and $\{ \tilde{\mathbf{y}}_{w}\}$ are used to create training datasets for the CNN classifier implemented at Bob. 

\subsection{Implementation of CNN Classifier}

Towards using a CNN, the collected sets of \{$\tilde{\mathbf{y}}_{s}$\} and \{$\tilde{\mathbf{y}}_{w}$\} (as explained in Section \ref{RFF_model}) are converted into scatter plots, and these figure files are saved and labelled as either ``legitimate" or ``adversary" depending on the frame. Henceforth, for the rest of the classification process, these figure files are the instances of the dataset used for training the CNN. After training, the goal of the CNN-based classifier is to identify whether Alice is the transmitter responsible for the received frame. Recall that the CNN model extracts relevant features and achieves accurate classifications through training on labeled datasets. Our  CNN architecture employs convolutional layers with filter sizes of $[3\times3]$ to capture local features from the scatter plots, using $256$, $128$, and $64$ image filters. Furthermore, rectified linear unit activation functions are used to enhance nonlinearity, and max-pooling layers are used to reduce input dimensionality while preserving crucial features. Finally, the fully connected layer uses a softmax activation function for image classification. For our experiments, we use learning rate of $0.001$ and batch size of $32$.


\section{Attack Detection via Federated Learning}\label{fl_inject}

Although a CNN based classifier can be implemented in the single base-station model, its detection accuracy is limited by the availability of the training datasets on both Alice's and Dave's samples. To further improve its accuracy, we apply FL under the framework of multiple base-stations, as shown in Fig. \ref{model}. In this context, we assume that multiple base-stations across various geographical areas experience the same class of injection attacks on some of its clients. As a result, the proposed secret-key based mitigation strategy is already implemented at the clients of each base-station along with the corresponding CNN based detection at the base-stations. Furthermore, we assume that the base-stations are connected to a server-station which is responsible to execute the following tasks: (i) to receive the weights of the CNN from each base-station, (ii) to execute the FedAvg algorithm \cite{s2}, and (iii) to distribute the averaged weights back to the base-stations. In the rest of this section,  we delve into FL's practical implementation and present its performance improvement over the single-base-station framework. 
\begin{figure}[ht]
		\centering
		\includegraphics[scale=0.25]{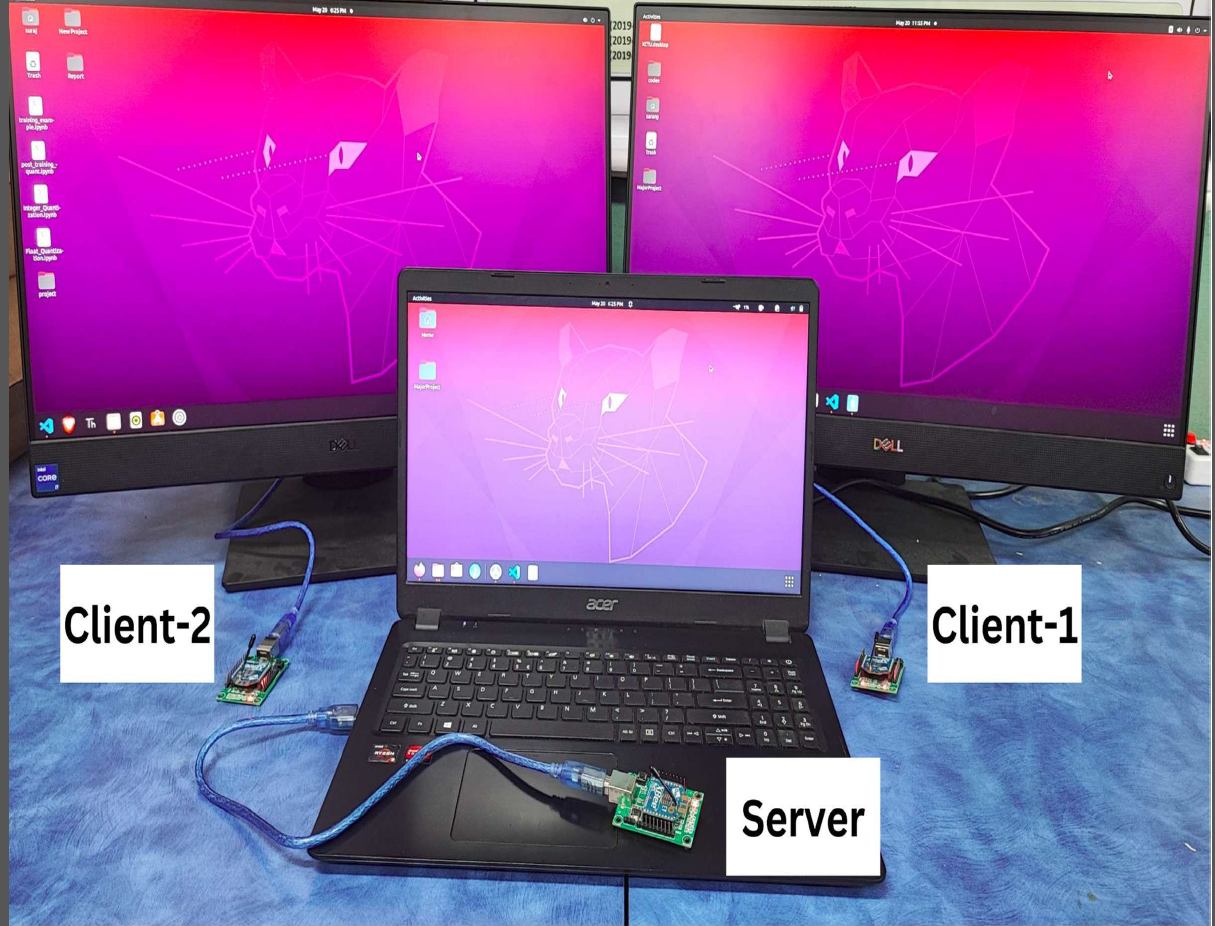}
		\caption{Depiction of FL prototype using XBee devices with two clients (base-stations), wherein the CNN model parameters are averaged over multiple iterations.}
		\label{xbee}
  \vspace{-4mm}
	\end{figure}

\subsection{Testbed Details for Federated Learning}
\label{back_results}

Our FL setup involves two models, one with two base-stations and a server-station, and the other with four base-stations and a server-station. In both these cases, XBee devices, as shown in Fig. \ref{xbee}, play the role of base-stations and server-station. We utilized the XBee S$2$C model, a member of the XBee Series $2$ family. This module functions on the IEEE $802.15.4$ wireless protocol, offering a low-power, short-range communication solution with support for radio-frequency data rates up to $250$ kbps. For more details on the specifications of the XBee setup, we refer the reader to Table \ref{back-table}. In our implementation, the server's XBee acts as the coordinator while the base-station XBees serve as routers. The coordinator initializes and maintains the network, with routers communicating solely with the coordinator. Each base-station independently trains its CNN, with communication facilitated through XBee devices. Base-stations train local models on individual datasets, and the server aggregates updates through federated averaging in multiple communication rounds, ensuring continual model improvement. Finally, in the evaluation phase, we test the updated model at each base-station with the dataset specified in earlier sections.

\begin{table}[!ht]
\centering
\addtolength{\tabcolsep}{-15.5pt}
\caption{\small{\textbf{Details XBee's configuration for FL}}}
\label{back-table}
\resizebox{7.5cm}{!}{%
\begin{tabular}{|c|c|}
\hline
\textbf{Hardware} & \textbf{Software} \\ \hline
$3$ XBee S$2$C devices. & TensorFlow v$2.11.0$ \\ \hline
\multirow{2}{*}{$3$ Ubuntu v$20.04.4$ LTS Systems (i$9$ $11$th gen.)} 
                 & digi-XBee v$1.4.1$ \\ \cline{2-2}
                 & Python and MATLAB \\
                \hline
\end{tabular}%
}
\end{table}

\begin{table*}
\centering
\vspace{0.6cm}
\addtolength{\tabcolsep}{-4.8pt}
\caption{\small{\textbf{FL setup attack detection results}}}
\label{gen_table}
\resizebox{14.8cm}{!}{%
\begin{tabular}{|c|c|c|c|c|c|c|c|}
\hline
\multirow{2}{*}{\textbf{Measure}} & \multicolumn{2}{|c|}{\textbf{Value (2-Client Setup)}} & \multicolumn{4}{|c|}{\textbf{Value (4-Client Setup)}} & {\textbf{Value}}\\  
\cline{2-8}
    & Client-1 & Client-2 &  Client-1 & Client-2 & Client-3 & Client-4 & Central Node \\
\hline
 Sensitivity & 0.9810 & 0.9900 & 0.9600 & 0.9680 & 0.9900 & 0.9850 & 0.9940 \\ \hline
 Specificity & 0.9990 & 0.9990 & 0.8600 & 0.8620 & 0.9990 & 0.8850 & 0.9660 \\ \hline
  Precision & 0.8926 & 0.8684 & 0.8727 & 0.8752  & 0.8684  & 0.8955 & 0.9669 \\ \hline
Accuracy & 0.9315 & 0.9200  & 0.9100 & 0.9150  & 0.9200  & 0.9350 & 0.9800 \\  \hline
F1 Score & 0.9347 & 0.9252  & 0.9143 & 0.9193  & 0.9252  & 0.9381 & 0.9803 \\ \hline
\end{tabular}%
}
\end{table*}


\subsection{Experimental Results on Federated Learning}
\label{fl_results}

\begin{figure}[ht!]
    \centering
    \includegraphics[scale = 0.35]{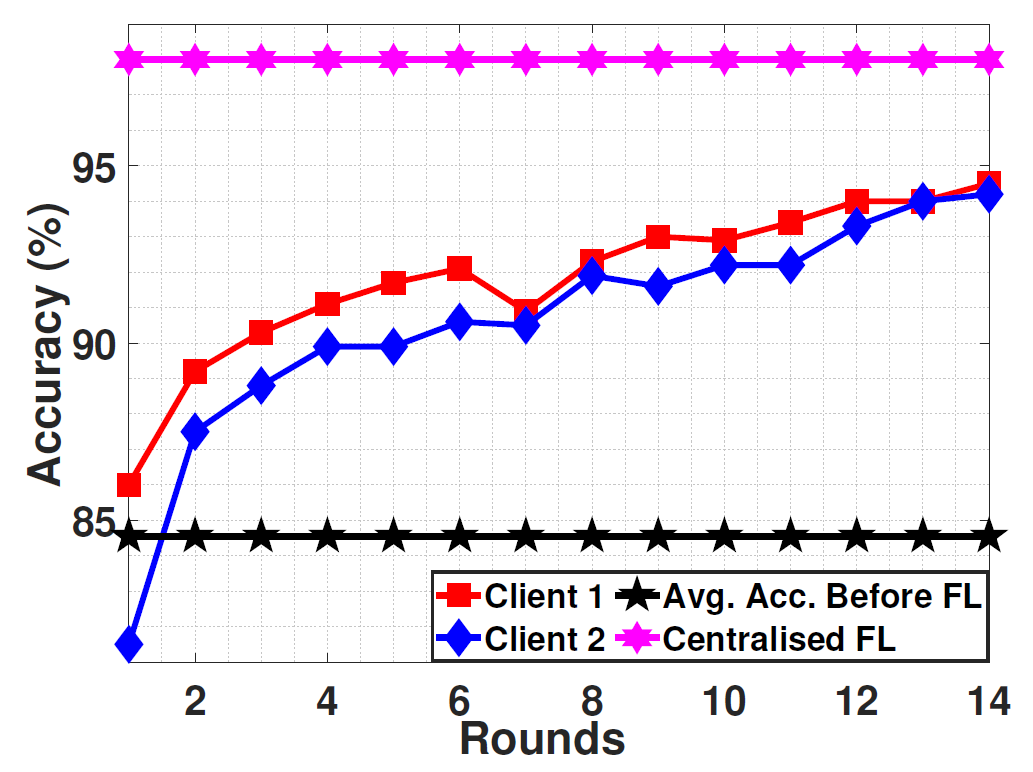}
    \caption{Accuracy improvements as a function of the number of rounds with two base-stations. In the legends of the figure, clients refer to the base-stations that are involved in FL.}
    \label{c1c2}
      \vspace{-1mm}
     \end{figure}

First, we present the experimental results for FL with $2$ base-stations. For this configuration, each base-station has $10,000$ scatter plots ($5,000$ per class) for training and $1,500$ ($750$ per class) for testing. With regards to the CNN at each base-station, the model comprises a total of 7253 parameters. Upon
extracting the weights from the model and saving them in a weight file, a file
with a size of 30 KB was generated for every round of FL interaction. As shown in Fig. \ref{c1c2}, after $10$ FL rounds, base-station 1 (referred to as client 1) achieves roughly  $93\%$ accuracy, and base-station 2 (referred to as client 2) approaches $92\%$, compared to an initial average accuracy of $84.55\%$ per client without FL. We also note that the centralized CNN yields $98\%$ accuracy, wherein a single CNN is trained and tested on the union of the dataset at the two base-stations. Additional insights are detailed in Table \ref{gen_table}. Due to the substantial size of the model parameters and the low-bandwidth of XBee devices, each round of FL consumed approximately 2.5 minutes, resulting in a
total duration of 25 minutes over 10 rounds. 

\begin{figure}[ht!]
    \centering
    \includegraphics[scale = 0.35]{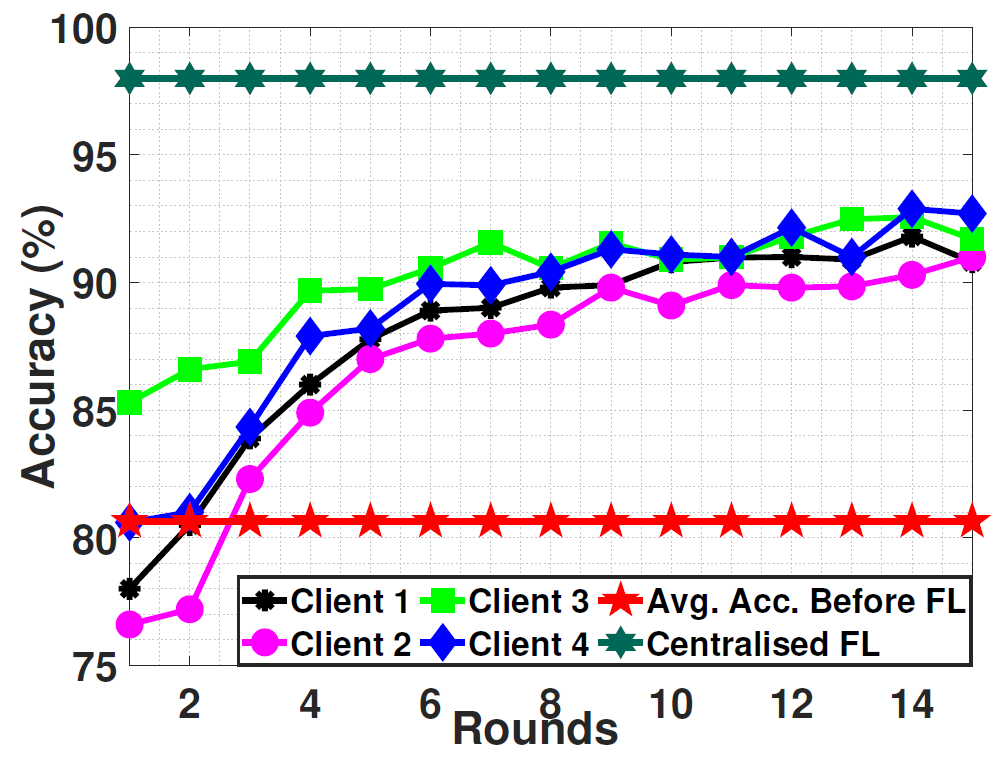}
    \caption{Accuracy improvements as a function of the number of rounds with four base-stations. In the legends of the figure, clients refer to the base-stations that are involved in the FL process.}
    \label{c1c4}
    \vspace{-1mm}
    \end{figure}

We also present the testbed results for 4 base-stations and one server-station. Each base-station has $5,000$ scatter plots ($2,500$ per class) for training and $750$ ($375$ per class) for testing. The CNN model size is same as that with two base-stations. Results in this configuration are illustrated in Fig. \ref{c1c4}, which reveal significant enhancements in overall accuracy for each base-station. After $15$ rounds of FL, observations indicate that base-station $1$ achieved around $91\%$ overall accuracy, base-station $2$ reached approximately $91.5\%$, and base-station $3$ and base-station $4$ attained about $92\%$ and $93.5\%$, respectively. In contrast, the average accuracy per base-station before implementing FL was $80.64\%$. Furthermore, testing with a centralized model, leveraging complete data, resulted in an accuracy rate of $98\%$, aligning with expectations based on the full dataset. Further insights on the results can be found in Table \ref{gen_table}. Similar to the case of two base-stations, in our setup, each round of FL consumed approximately 5 minutes due to low-bandwidth of XBee networks, along with the fact that each base-station communicated with the server-station in time-division multiple access manner to communicate their weights. While we understand that the presented orders of delay to execute FL are unacceptable in practice, using high-rate backhaul networks can speed-up this process in the order of few milliseconds in the real world setting.  

\section{Conclusion}\label{concl}

We have developed a prototype for detecting injection attacks in wireless networks. We have used Adalm Pluto based SDRs to implement a secret-key based communication strategy, which was designed to distinguish the legitimate node from an adversary through the scatter plots of the received I/Q samples. Our results demonstrate that an FL-based CNN classifier, when implemented at the base-station, can effectively detect injection attacks with high-accuracy.

\bibliographystyle{splncs04}

\begin{thebibliography}{16}

\bibitem{a30}
Ammar, Mahmoud, Giovanni Russello, and Bruno Crispo, ``Internet of Things: A survey on the security of IoT frameworks," \emph{Journal of Information Security and Applications}, vol. 38, pp. 8-27, 2018.

\bibitem{a29}
Alromih, Arwa, Mznah Al-Rodhaan, and Yuan Tian,``A randomized watermarking technique for detecting malicious data injection attacks in heterogeneous wireless sensor networks for internet of things applications," \emph{Sensors}, 18.12 (2018): 4346.	

\bibitem{a2} I. O. Kennedy, P. Scanlon, F. J. Mullany, M. M. Buddhikot, K. E. Nolan, and T. W. Rondeau, “Radio transmitter fingerprinting: A steady state frequency domain approach,” \emph{IEEE 68th Vehicular Technology Conference}, pp. 1-5, 2008.

\bibitem{a4} S. U. Rehman, K. W. Sowerby, S. Alam, I. T. Ardekani, and D. Komosny, “Effect of channel impairments on radiometric fingerprinting,” in 2015 \emph{IEEE International Symposium on Signal Processing and Information Technology (ISSPIT)}, pp. 415-420, 2015.

\bibitem{s110}
Albers, Patrick, et al. ``Security in Ad Hoc Networks: a General Intrusion
Detection Architecture Enhancing Trust Based Approaches,” \emph{Wireless
Information Systems}, 2002.

\bibitem{a5}  K. Youssef, L.-S. Bouchard, K. Z. Haigh, H. Krovi, J. Silovsky, and C. P. V. Valk, “Machine learning approach to RF transmitter identification,” in \emph{IEEE Journal of Radio Frequency Identification}, vol. 2, no. 4, pp. 197-205, Dec. 2018.

\bibitem{s1}
Zhu, Guangxu et al.``Toward an intelligent edge: Wireless communication meets machine learning," \emph{IEEE communications magazine}, vol. 58, no. 1, pp. 19-25, Jan. 2020.

\bibitem{s3}
Kairouz, Peter et al. ``Advances and open problems in federated learning," \emph{Foundations and Trends in Machine Learning}, 2021.

\bibitem{s2}
Konečný, Jakub et al.``Federated optimization: Distributed machine learning for on-device intelligence," arXiv preprint arXiv:1610.02527, 2016.

\bibitem{s4}
N. Goel, V. Chaudhary and J. Harshan, ``Secret-key based non-coherent signalling to mitigate reactive injection attacks," \emph{IEEE International Conference on Signal Processing and Communications (SPCOM)}, pp. 1-5, 2022.

\end{thebibliography}

\end{document}